\documentstyle[aps,prl,epsfig]{revtex}
\begin{document}
\tighten
\twocolumn[\hsize\textwidth\columnwidth\hsize  
\csname @twocolumnfalse\endcsname              

\title{Inclusive and Direct Photons in
S + Au Central Collisions at 200A GeV/c}

\vspace{0.1in}

\author{Wang Hui,$^{2}$ \,\, Sa Ben-Hao,$^{1,2,3}$
 \,\, Tai An,$^{1,4}$ \,\,  and Sun Zu-Xun$^{2}$}

\vspace{0.1in}

\address{
$^1$ CCAST (World Lab.), P.O. Box 8730 Beijing, China \\
$^2$ China Institute of Atomic Energy, P.O. Box 275 (18),
     Beijing, 102413 China \footnotemark \\
$^3$ Institute of Theoretical Physics, Academic Sinica,
     Beijing, China\\
$^4$ Dept. of Physics and Astronomy, Univ. of California at Los Angeles,
     Los Angeles, CA 90025
}

\vspace{0.1in}

\date{\today}

\maketitle

\begin{abstract}
A hadron and string cascade model, JPCIAE, which is based on LUND
string model, PYTHIA event generator especially, is used to study
both inclusive photon production and direct photon production in
200A GeV S + Au central collisions. The model takes into account
the photon production from the partonic QCD scattering process,
the hadronic final-state interaction, and the hadronic decay and
deals with them consistently.  The results of JPCIAE model reproduce
successfully both the WA93 data of low p$_T$ inclusive photon
distribution and the WA80 data of transverse momentum dependent
upper limit of direct photon. The photon production from different
decay channels is investigated for both direct and inclusive photons.
We have discussed the effects of the partonic QCD scattering and the
hadronic final-state interaction on direct photon production as
well.

\vspace{0.15in}
\end{abstract}


\vspace{0.4in}

   ]  

\narrowtext

\footnotetext{\footnotemark Mailing address. E-mail:
hui@iris.ciae.ac.cn}

\section{Introduction}

Currently, relativistic nucleus-nucleus collisions are used to
study the characteristics of nuclear matter under extreme
conditions. One of the goals in this study is to search for the
quark-gluon plasma (QGP), which is believed to be formed at high
temperature and/or density \cite{Sta92,Har96,QM96,Hwa95} during a
nucleus-nucleus collision at high energies. Photons arising from
the electromagnetic interactions of the constituents of the plasma
will provide information on the properties of the plasma at the
time of their production \cite{Shu80}. Since photons hardly
reinteract in the produced medium, they form a relatively `clean'
probe to study a  QGP state. The possible detection, in near
future, of the photon produced in the QGP phase in relativistic
heavy-ion collider (RHIC) and/or large hadron collider (LHC) will
be of great interest in probing such a QGP state, but presently
that might not be the case at CERN SPS energies
\cite{Hwa85,McL85,Kaj86,Kap91,Sri92,WA80,WA98,WA9897,Xio92,Hun96}.

Photons measured after the subtraction of the photons from meson
decays are usually called ``direct photon''. The direct photons
could be produced from the interaction of matter in the QGP phase,
a mixed QGP and hadron phase, and a pure hadron phase
\cite{Sta92,Har96}. The thermal direct photon and prompt direct
photon are referred to the photons produced from the partonic QCD
processes in the QGP phase and in the hadron phase, respectively.
However, the photons produced in the hadronic interactions are
sorted into hadronic direct photon, although the prompt direct
photon is also originated from hadrons. Different processes give
rise to photons in different (transverse) momentum regions. Of
course, photons, which might show up in the low transverse
momentum region, extending to the region of intermediate
transverse momentum of 1 $-$3 GeV/c, from a QGP are specially
important. Photons with energies up to 2 GeV can come from the
decay of $\pi^0$ and $\eta$ resonances as well as from $\rho,
\omega, \eta'$, and $a_1$, and from the interaction of hadron
matter via $\pi \rho \rightarrow \gamma \pi$ and $\pi\pi
\rightarrow \gamma \rho$ reactions (i.e., hadronic direct photons)
\cite{Xio92,Hun96,Ste96,Li97}. If a hot quark-gluon plasma is
formed initially, clear signal of photon from the plasma could be
visible by examining photons with $p_T$ in the region of 2$-$3
GeV/c \cite{Kaj86,Kap91,Sri92}. However, photons in this
transverse momentum region could also be produced in the collision
of a parton from the projectile nucleon with another parton from
the target nucleon, i.e., prompt direct photon. Such a
contribution must be subtracted in order to infer the net photons
from the QGP source, i.e., thermal direct photon.

WA80 experimental data of transverse momentum dependent upper
limits of direct photon in 200A GeV/c S + Au central
collisions \cite{WA80} initiate theoretical interests extensively
\cite{Hun96,Li97,Sri94,Dum94,Dum98,Son98,Sol97}. Refs.
\cite{Sri94} and \cite{Dum94} concluded that WA80 data can only be
understood if a scenario with QGP phase transition is assumed.
However, in Ref. \cite{Li97} the single photon production has been
calculated using relativistic hadronic transport model and taking
into account self-consistently the change of hadron mass in dense
matter. It was found there that the spectra with either free or
in-medium meson mass do not exceed the WA80 upper limits, although
the experimental transverse momentum distribution in low $p_T$
region was not reproduced quite well. Ref. \cite{Hun96} calculated
the direct photon production using the rate theory (thermal model)
and the hydrodynamical model for space-time evolution of
temperature, their results were below WA80 upper limits.

Recently, WA93 measured the invariant differential
cross section distribution of inclusive photons at low transverse
momentum in S + Au central collisions at 200A GeV/c \cite{WA93}.
The results indicated that the photon yields at low transverse
momentum are much enhanced in comparing with the results of
photons from hadron decays measured by WA80 and with the VENUS
4.12 calculations.

We have already studied in details the photon production from QCD
hard processes in high transverse momentum region in
nucleon-nucleon, nucleon-nucleus, and nucleus-nucleus collisions
with the effects of parton intrinsic transverse momentum and the
contributions of next-to-leading-order Feynman diagram corrections
\cite{Won98}. The results show that the inclusion of intrinsic
transverse momentum of parton leads to an enhancement of photon
production cross section and the enhancement increases as
$\sqrt{s}$ decrease. Such an enhancement is an important
consideration in the region of photon momenta under investigation
in high energy heavy-ion collisions.

In this paper, we study the low transverse momentum distribution
of both inclusive photons and direct photons produced in S + Au
central collisions at 200A GeV/c using a hadron and string
cascade model, JPCIAE \cite{Sa99,Sa99b}. We have considered
consistently the photons from different sources, such as the
partonic QCD scattering, the hadronic final-state interaction, and
the hadron decay. Our results reproduce successfully both the WA80
data of direct photon upper limits and the WA93 data of low
transverse momentum inclusive photons. We have also compared the
contributions from different hadron decays in inclusive photon
production. The effects of the partonic QCD processes and the
hadronic final-state interactions on direct photon production are
discussed as well.

\section{BRIEF DESCRIPTION FOR THE MODEL}

A hadron and string cascade model, JPCIAE, was proposed to
describe the relativistic nuclear collision \cite{Sa99,Sa99b}. In
JPCIAE the simulation is performed in the laboratory system. The
origin of coordinate space is positioned at the center of the
target nucleus and the beam direction is taken as the $z$ axis. As
for the origin of time it is set at the moment when the distance
between the projectile and target nuclei along z direction is
equal to zero (the collision time can be negative).

A colliding nucleus is depicted as a sphere with radius $\sim$
1.05 A$^{1/3}$ ($A$ refers to the atomic mass number of this
nucleus) in its rest frame. The spatial distribution of nucleons
in this frame is sampled randomly due to the Woods-Saxon
distribution. The projectile nucleons are assumed to have an
incident momentum and the target nucleons are at rest. That means
the Fermi motion in a nucleus and the mean field of a nuclear
system are here neglected due to relativistic energy in question.
For the spatial distribution of the projectile nucleons the
Lorentz contraction is taken into account. A formation time is
given to each particle and a particle starts to scatter with
others after it is ``born''. The formation time is a sensitive
parameter in this model, see Ref. \cite{Sa99} for the details.
\begin{figure}[htbp]
 \centerline{
            \epsfig{file=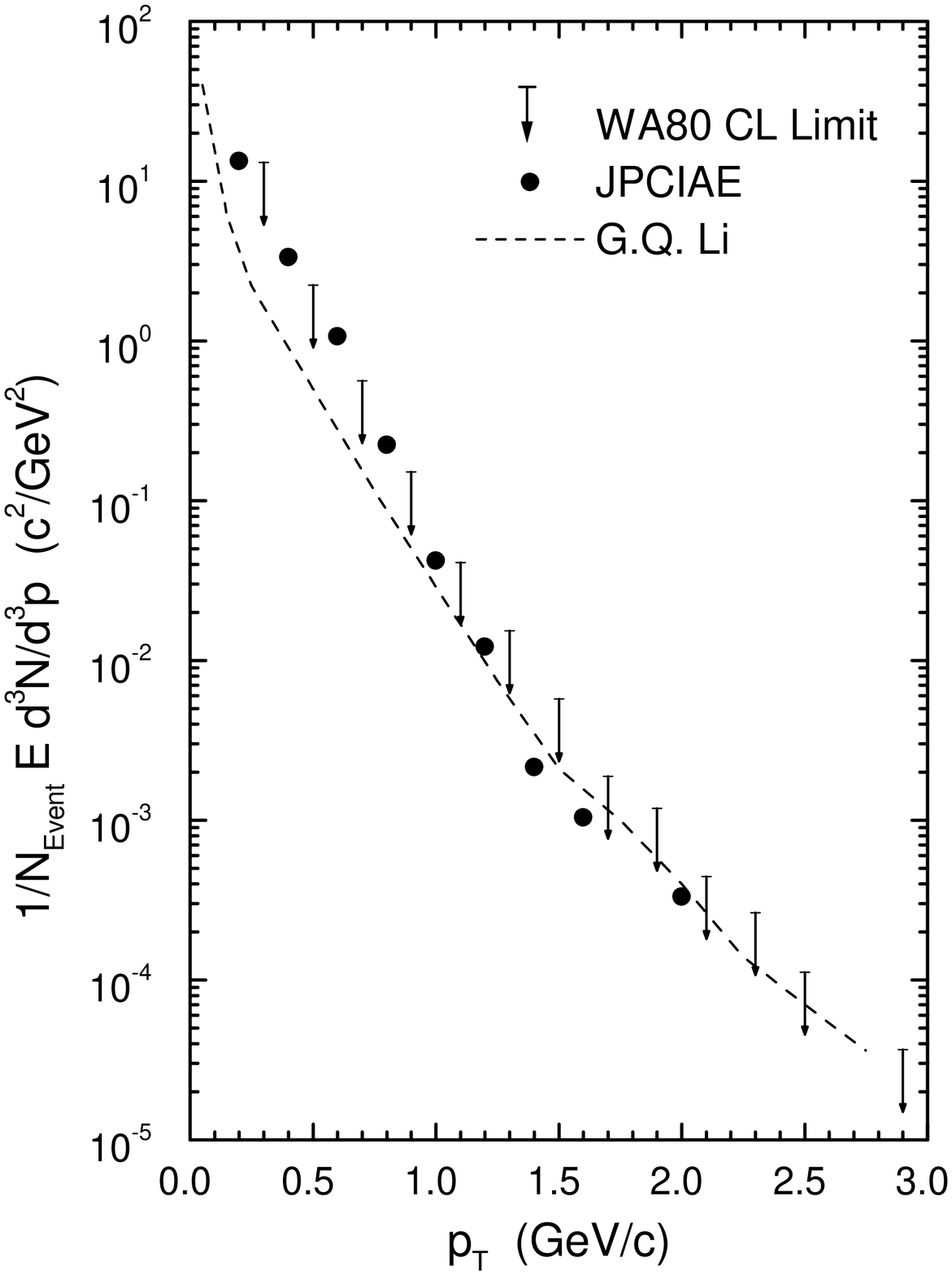,width=8cm}
 }
  \caption[wa80]{
Transverse momentum distribution of direct photon (in rapidity
range of 2.1 $\leq$y$\leq$ 2.9) produced in S + Au central
collisions at 200A GeV/c. The arrows stand for WA80 upper limits
at the 90\% confidence level \cite{WA80}, the solid circles refer
to the results of JPCIAE, and the dashed curve are the results of
\cite{Li97}. \label{wa80}}
\end{figure}

A collision time is calculated according to the requirement that
the minimum approaching distance of a colliding pair should be
less than or equal to the value $\sqrt{\sigma_{tot}/\pi}$, where
$\sigma_{tot}$ is the total cross section of the colliding pair.
The minimum distance is calculated in the center of mass system
(C.M.S.) frame of the two colliding particles. If these two
particles are moving towards each other at the time when both of
them are ``born'', the minimum distance is defined as the distance
perpendicular to the momenta of both particles. If the two
particles are moving back to back, the minimum distance is defined
as the distance at the moment when both of them are ``born''. All
the possible collision pairs are then ordered into a collision
time sequence, called the collision time list. The initial
collision time list is composed of the colliding nucleon pairs, in
each pair here one partner is from the projectile nucleus and the
other from the target nucleus.
\begin{figure}[htbp]
 \centerline{
            \epsfig{file=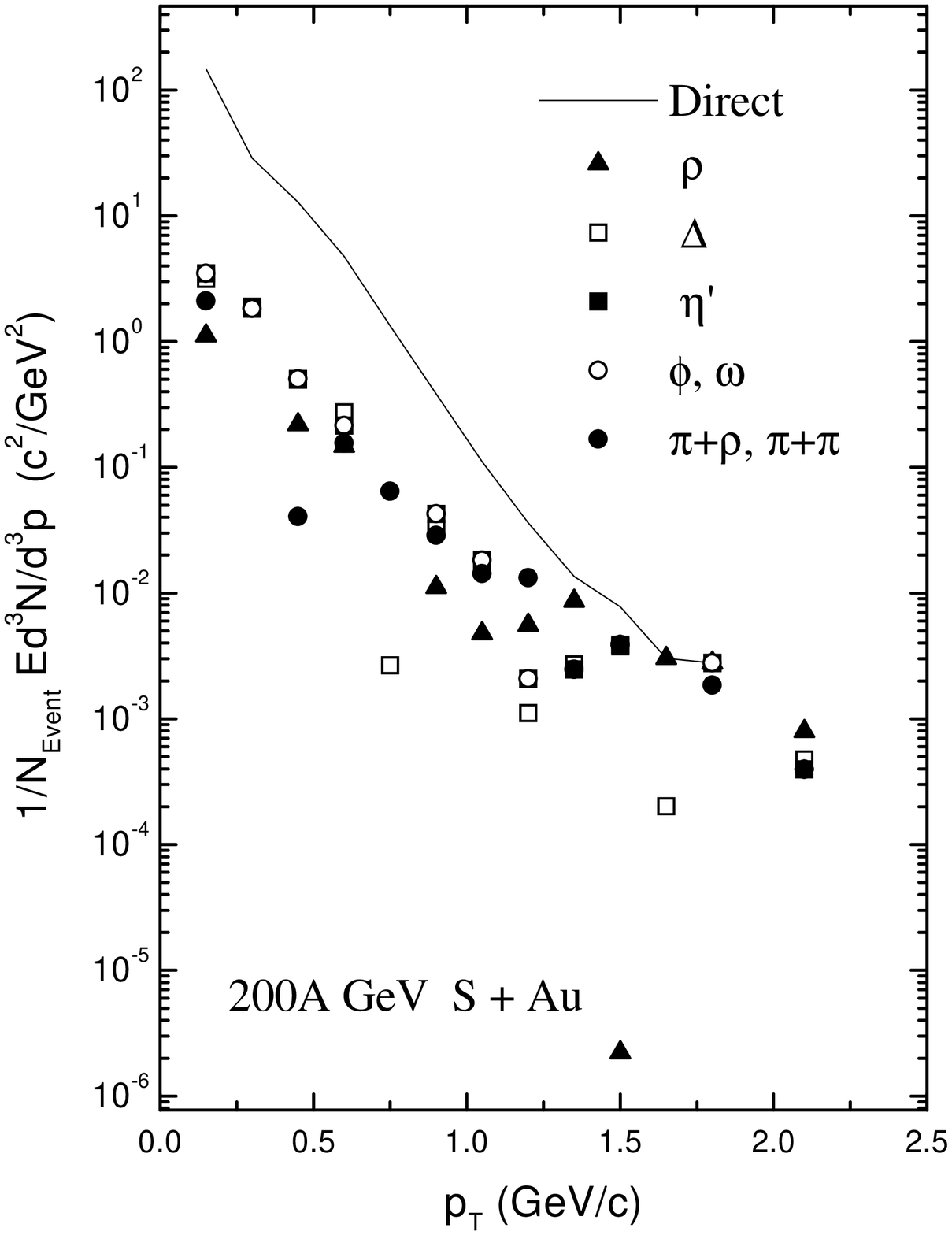,width=8cm}
 }
  \caption[wa802]{
The solid curve is the theoretical transverse momentum distribution
of direct photons (i.e. the inclusive photons after subtraction of
$\pi^0$ and $\eta$ decay photons) in full rapidity space produced
in S + Au central collisions at 200A GeV/c. The full triangles,
the open squares, the full squares, and the open circles are the
$\rho$, $\Delta$, $\eta'$, and $\phi$ and $\omega$ decay photons,
respectively. The photons from hadronic interactions, cf. Eqs.
\ref{hh}, and \ref{hh1} are shown by full circles.
\label{wa802}}
\end{figure}

Then the pair with the least collision time in the initial
collision time list is selected to start the first collision. If
the c.m.s. energy, $\sqrt{s}$, of this colliding pair (a hadron-
hadron collision) is larger than or equal to $\sim$ 4 GeV, two
string states are formed and PYTHIA is called to produce the final
state hadrons (scattered state). Otherwise no string state is
formed and the conventional two-body scattering process
\cite{Cu81,Be88,Sa98} is executed. After the scattering of this
colliding pair, both the particle list and the collision time list
are then updated and they are now not only composed of the
projectile and target nucleons but also the produced hadrons.
Repeat the previous steps to perform the second collision, the
third collision, $\cdots$, until the collision time list is empty,
i.e. no more collision occurs in the system. Finally, we consider
the decay of the unstable particles.

In PYTHIA we consider not only the low-$p_T$ processes but also
the high-$p_T$ processes of the particle production
\cite{Sj94,And81}. Many partonic QCD scattering processes
including photon production have been considered. A user is
allowed to run the program with any desired subset of those
processes.
\begin{figure}[htbp]
 \centerline{
            \epsfig{file=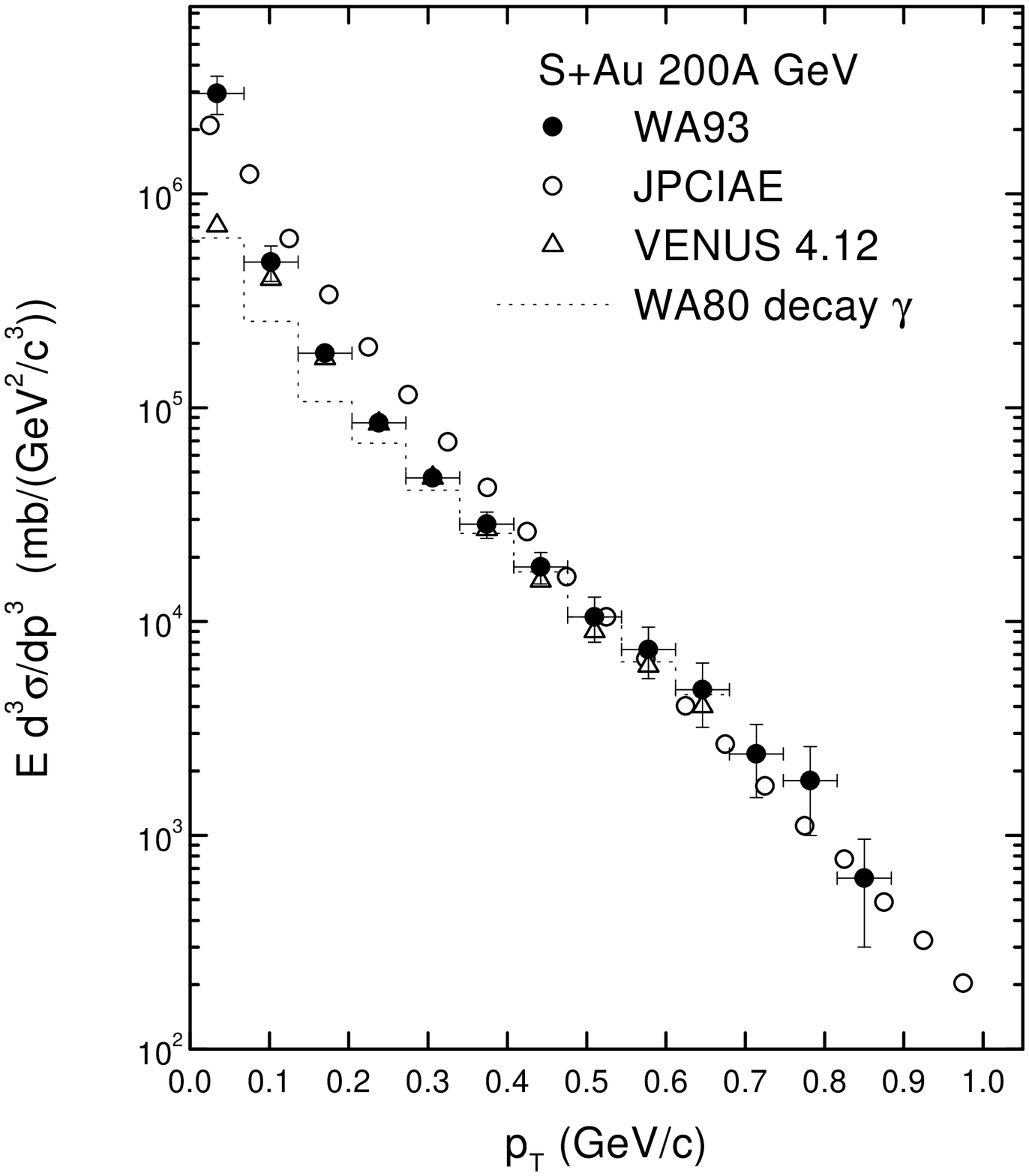,width=8cm}
 }
  \caption[]{
 The transverse momentum distribution of inclusive photons in
 S + Au central collisions at 200A GeV/c. The solid circles
 represent WA93 data, the open circles refers to the results of
 JPCIAE normalized to the experimental data at $p_T$ = 0.5 GeV/c,
 the open triangles are the results of VENUS 4.12 normalized
 to the experimental data at $p_T$ = 0.3 GeV/c and the dotted
 histogram is WA80 decay photons.
\label{wa93}}
\end{figure}

We have inspected this model and the corresponding event
generator, JPCIAE, by comparing model predictions with the NA35
data of the charge multiplicity, the rapidity and the transverse
momentum distributions of the negative charge particles (h$^-$)
and the participant protons in $p p$, $p A$, and $A B$ collisions
\cite{Sa99}. The agreements between theory and experiment are
reasonably good. The model has explained successfully the $J/\psi$
suppression in $p A$ and $A B$ (including $Pb Pb$) as well
\cite{Sa99,Sa99b}.

\section{RESULTS AND DISCUSSIONS}

We have studied the photon production in S + Au central collisions
at 200A GeV/c using JPCIAE model. Following partonic QCD
scattering processes with photon emission were selected
\begin{eqnarray}\label{py}
  f_i + \bar{f}_i &\rightarrow & g + \gamma ,\\
  f_i + \bar{f}_i &\rightarrow & \gamma + \gamma ,\\
  f_i + g &\rightarrow & f_i + \gamma ,\\
  g   + g &\rightarrow & \gamma + \gamma ,\\
  g   + g &\rightarrow & g + \gamma ,
\end{eqnarray}
where $f_i$ refers to the quark with $i$ flavor and both
low-$p_T$ and high-$p_T$ contributions are included. The
hadronic photon production reactions
\begin{equation}\label{hh}
  \pi + \pi \rightarrow \rho + \gamma ,
\end{equation}
and
\begin{equation}\label{hh1}
  \pi + \rho \rightarrow \pi + \gamma ,
\end{equation}
were taken into account as well. For simplicity, the isospin
averaged parameterization formulas \cite{Kap91,Li97} were used for
the relevant cross sections here. Of course, the hadron decays,
such as $\pi^0$, $\eta$, $\rho$, $\omega$, $\eta'$, $a_1$,
$\Delta$, etc. were included as well.

The results of the transverse momentum distribution of direct
photons (i.e., the inclusive photons after subtraction of $\pi^0$
and $\eta$ decay photons) produced in a rapidity range of 2.1
$\leq$ y $\leq$ 2.9 in central S + Au collisions at 200A GeV/c are
given in Figure \ref{wa80}. In this figure the arrows stand for WA80
upper limits at the 90\% confidence level \cite{WA80}, the solid
circles refer to the results of JPCIAE calculated with the same
centrality and rapidity cuts as WA80, and the dashed curve represents
results of \cite{Li97}. One sees from this figure that in comparing
with the WA80 data the results of JPCIAE is somewhat better than
Ref.\cite{Li97}, in low $p_T$ region especially. That might be
attributed to the contributions from partonic QCD processes
(low-$p_T$ processes) and from emission of photons off quarks and
leptons in shower. In addition, JPCIAE model is also more
self-consistent than one used in \cite{Li97} where the results of
RQMD has to be an input of hadronic transport simulation.

The JPCIAE results of transverse momentum distribution of direct
photons in full rapidity space produced in central S + Au collisions
at 200A GeV/c are given in Figure \ref{wa802} as solid curve.
In this figure the full triangles, the open squares, the
full squares, and the open circles are the $\rho$, $\Delta$, $\eta'$,
and $\phi$ and $\omega$ decay photons, respectively. The photons
from hadronic interactions, cf. Eqs. \ref{hh}, and \ref{hh1} , are
shown by full circles. From Fig. \ref{wa802} we can see that at
low transverse momentum region the sum of above decay photons and
the photons from hadronic interactions is far below the direct
photons. That indicates the prompt direct photon is visible at
low transverse momentum region in the case of no QGP formation.
\begin{figure}[htbp]
 \centerline{
            \epsfig{file=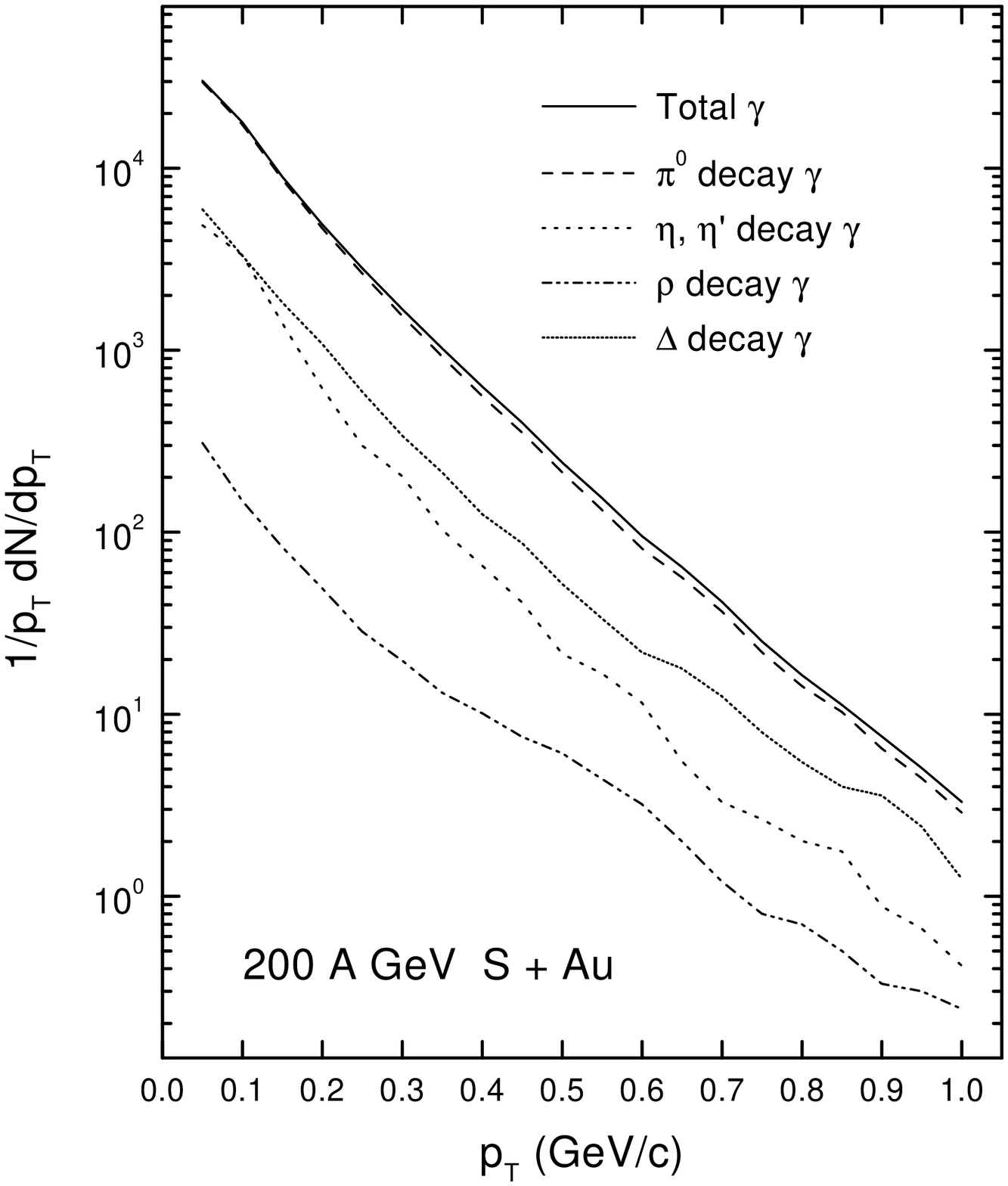,width=8cm}
 }
  \caption[wa932]{
The transverse momentum distributions of photons in central
collisions S + Au at 200A GeV/c. The solid, the long-dashed, the
dotted, the dashed, and the dot-dashed curves represent the
inclusive photons, the $\pi^0$ decay, the $\Delta$ decay, the
$\eta$ and $\eta'$ decays, and the $\rho$ decay photons,
respectively. \label{wa932}}
\end{figure}

After WA80 measured the upper limits of direct photon in S + Au
central collisions at 200A GeV/c, WA93 measured further the
low transverse momentum distribution of inclusive photons for the
same reaction system. WA93 data (solid circles with error bar)
are compared with the WA80 results of decay photons (dotted
histogram) and with the theoretical calculations of VENUS 4.12
(open triangles, normalized to the experimental data at $p_T$ =
0.3 GeV/c)) in Figure \ref{wa93}. In this figure the open circle
are the results of JPCIAE normalized to the experimental data at
$p_T$ = 0.5 GeV/c. From this figure one knows that the WA93 datum
point below $p_T$ = 0.1 GeV/c is distinctly larger than the WA80
decay photons and the VENUS 4.12 results. However, JPCIAE
reproduces this WA93 datum point and others successfully due
to the contributions from partonic QCD processes (low-$p_T$
processes) and from emission of photons off quarks and leptons
in shower were taken into account in JPCIAE calculations.

We have calculated also the transverse momentum distributions of
different decay photons in S + Au central collisions at 200A GeV
/c using JPCIAE model, as shown in Fig. \ref{wa932}. In this
figure the solid, the long-dashed, the dotted, the dashed, and the
dot-dashed curves represent inclusive photons, the $\pi^0$ decay,
the $\Delta$ decay, the $\eta$ and $\eta'$ decay, and the $\rho$
decay photons, respectively. The results indicate that the $\pi^0$
decay photons are absolutely dominant in the production of
inclusive photons. The contributions from $\eta$, $\eta'$, and
$\rho$ decay and even from $\Delta$ decay are much less than one
from the $\pi^0$ decay, which accounts for nearly 98.8 $\%$ of
total decay photons.
\begin{figure}[htbp]
 \centerline{
            \epsfig{file=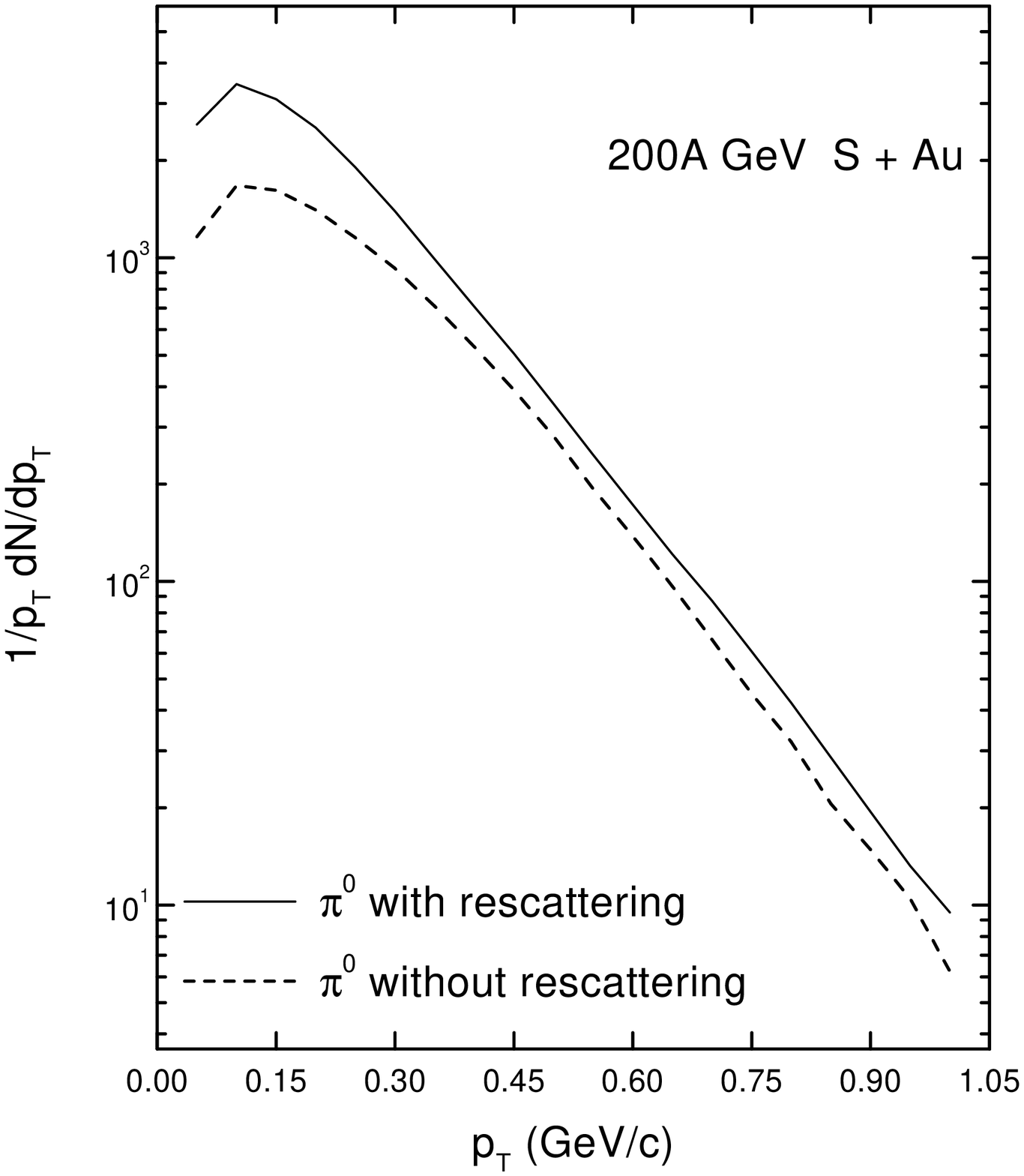,width=8cm}
 }
  \caption[pi0]{
The rescattering effects to the $\pi^0$ production in central S +
Au collisions at 200A GeV/c. The solid and the dashed curves are
the results with and without rescattering, respectively.
\label{pi0}}
\end{figure}

Figure \ref{pi0} shows the rescattering effect to the $\pi^0$
production in S + Au central collisions at 200A GeV/c. In this
figure the transverse momentum distributions of $\pi^0$ with and
without rescattering are shown, respectively, by the solid and the
dashed curves. The total neutral pion yield with rescattering is
nearly twice of the one without rescattering. In the high
transverse momentum region the effect of rescattering leads
$\pi^0$ yield increasing nearly a factor of 1.5 . This increasing
factor is even bigger in the low transverse momentum region. The
$\pi^0$ yield with rescattering at $p_T$ = 0.05 GeV/c is about 2.4
times of the one without rescattering. The enhancement of $\pi^0$
yield due to rescattering also responds to the inclusive photon
production enhancement at low transverse momentum.

In summary, we have used the hadron and string cascade model,
JPCIAE, to study the photon production in 200A GeV/c S + Au
central collisions. The model takes into account photon production
from the partonic QCD process, the hadronic final-state
interaction, and the hadron decay and deals with them consistently.
The results of JPCIAE reproduce successfully both the WA80 data of
transverse momentum dependent upper limits of the direct photon
and the WA93 data of inclusive photon low transverse momentum
distribution. We have compared the contributions from different
hadron decays and the results show that $\pi^0$ is much dominant
in the inclusive photon production. Therefore $\pi^0$
rescattering plays an important role in low transverse momentum
distribution of inclusive photon. The partonic QCD processes
(low-$p_T$ processes) and the photon emission off quarks and leptons
in shower are quite important in transverse momentum distribution
at low $p_T$ region both for inclusive photon and direct photon.

\section{ACKNOWLEDGMENTS}
We would like to thank G. Q. Li for supply us the isospin averaged
parameterization formulas of hadronic direct photon production cross
sections and for discussions. Many thanks to T. Sj\"{o}strand for
detailed instructions in using PYTHIA. This work was supported by
national Natural Science Foundation of China and Nuclear Industry
Foundation of China.


\begin{thebibliography}{99}
\bibitem{Sta92}
J. Stachel and G. R. Young,
Annu. Rev. Nucl. Part. Sci. {\bf 42}, 537 (1992).

\bibitem{Har96} J. W. Harris and B. M\"uller,
Ann. Rev. Nucl. Part. Phys. {\bf 46}, 71 (1996).

\bibitem{QM96} Proceedings of the Quark Matter '96 Conference,
Heidelberg, 1996, edited by P. Braun-Munzinger, H. J. Specht, R.
Stock, and H. St\"ocker, published in Nuclear Physics, Vol. {\bf
A610} (1996).

\bibitem{Hwa95} R. C. Hwa, {\it Quark-Gluon Plasma}, Vol. 2, World
Scientific Publishing Company, 1995.

\bibitem{Shu80} E. V. Shuryak, Phys. Rep. {\bf 61}, 71 (1980);
L. McLerran, Rev. Mod. Phys. {\bf 58}, 1001 (1986)

\bibitem{Hwa85} R. C. Hwa and K. Kajantie, Phys. Rev. {\bf D32},
1109 (1985).

\bibitem{McL85} L. D. McLerran and T. Toimela, Phys. Rev. {\bf D31},
545 (1985).

\bibitem{Kaj86}
K. Kajantie, J. Kapusta, L. McLerran, and A. Mekjian, Phys. Rev.
{\bf D34}, 2746 (1986).

\bibitem{Kap91} J. Kapusta, P. Lichard, and D. Seibert,
Phys. Rev. {\bf D44}, 2774 (1991), [Erratum: {\it ibid.} {\bf
D47}, 4323 (1993).]

\bibitem{Sri92} D. K. Srivastava, B. Sinha, M. Gyulassy, and
X. N. Wang, Phys. Lett. {\bf B276}, 285 (1992).

\bibitem{WA80}
R. Albrecht $et~al.$ (WA80 Collaboration),
Phys. Rev. Lett. {\bf 76}, 3506 (1996).

\bibitem{WA98}
T. Peitzmann $et~al.$ (WA98 Collaboration),
invited talk presented at Quark Matter '96, Heidelberg, Germany,
May 20-24, 1996.

\bibitem{WA9897} B. Wyslouch $et~al.$, WA98 Collaboration, Proceedings
of Quark Matter '98, Japan, 1997.

\bibitem{Xio92} L. Xiong, E. V. Shuryak, and G. E. Brown,
Phys. Rev. {\bf D46}, 3798 (1992).

\bibitem{Hun96}
C. M. Hung and E. V. Shuryak,
Phys. Rev. {\bf C 56}, 453 (1997).

\bibitem{Ste96}
J. V. Steele, H. Yamagishi, and I. Zahed, Phys. Lett. {\bf B384},
255 (1996).

\bibitem{Li97}
G.Q. Li, G.E. Brown, C. Gale, and C.M. Ko, nucl-th/9712048.

\bibitem{Sri94} D.K. Srivastava and B. Sinha, Phys. Rev. Lett. {\bf
59}, (1994) 2421.

\bibitem{Dum94} A. Dumitru, U. Katscher, J. A. Maruhn,
H. St\"{o}cker, W. Greinre, and D. H. Rischke, Phys. Rev. {\bf C
51}, 2166 (1995).

\bibitem{Dum98}
A. Dumitru, M. Bleicher, S.A. Bass, C. Spieles, L. Neise, H.
St\"{o}cker, and W. Greiner, Phys. Rev. {\bf C 57}, 3271 (1998).

\bibitem{Son98}
C. Song and G. Fai, Phys. Rev. {\bf C 58}, 1689 (1998).

\bibitem{Sol97} J. Sollfrank, P. Huovinen, M. Kataja, P. V.
Ruuskanen, M. Prakash, and R. Venugopalan, Phys. Rev. {\bf C 55},
392 (1997).

\bibitem{WA93}
M. M. Aggarwal $et~al.$ (WA93 Collaboration), Phys. Rev. {\bf C
56}, 1160 (1997).

\bibitem{Won98}
Cheuk-Yin Wong and Hui Wang, Phys. Rev. {\bf C
58}, 376 (1998).

\bibitem{Sa99}
Sa Ben-Hao, Tai An, Wang Hui and Liu Feng-He, nucl-th/9803033 and
Phys. Rev. {\bf C} in press.

\bibitem{Sa99b}  Sa Ben-Hao, Amand Faessler, Tai An,
T. Waindzoch, C. Fuchs, Z.S. Wang, Wang Hui, nucl-th/9809020 and
J. Phys. G in press.

\bibitem{Cu81} J. Cugnon, T. Mizutani, J. Vandermeulen, Nucl.
Phys. {\bf A 352}, 505 (1981).

\bibitem{Be88} G. F. Bertsch, S. Das Gupta, Phys. Reports
{\bf 160}, 189 (1988).

\bibitem{Sa98} Sa Ben-Hao, Tai An, Comp. Phys. Commu. {\bf
90}, 121 (1995); $ibid.$ {\bf 116}, 353 (1999).

\bibitem{Sj94} T. Sj\"{o}strand, Comp. Phys. Commu. {\bf
82}, 74 (1994).

\bibitem{And81} B. Andersson, G. Gustafson, I. Holgersson and O.
M{\aa}nsson, Nucl. Phys. {\bf B 178}, 242 (1981).
\end{thebibliography}
\end{document}